# Digital confocal microscopy through a multimode fiber

**Damien Loterie,[1,*] Salma Farahi,[1,2] Ioannis Papadopoulos,[2] Alexandre Goy,[2,3] Demetri Psaltis,[2] and Christophe Moser[1]**

[1] *Laboratory of Applied Photonics Devices, School of Engineering, École polytechnique fédérale de Lausanne, CH-1015 Lausanne, Switzerland*

[2] *Laboratory of Optics, School of Engineering, École polytechnique fédérale de Lausanne, CH-1015 Lausanne, Switzerland*

[3] *Imaging Physics Group, Department of Electrical Engineering, Princeton University, Princeton, New Jersey 08544, USA*

[*]*damien.loterie@epfl.ch*

**Abstract:** Acquiring high-contrast optical images deep inside biological tissues is still a challenging problem. Confocal microscopy is an important tool for biomedical imaging since it improves image quality by rejecting background signals. However, it suffers from low sensitivity in deep tissues due to light scattering. Recently, multimode fibers have provided a new paradigm for minimally invasive endoscopic imaging by controlling light propagation through them. Here we introduce a combined imaging technique where confocal images are acquired through a multimode fiber. We achieve this by digitally engineering the excitation wavefront and then applying a virtual digital pinhole on the collected signal. In this way, we are able to acquire images through the fiber with significantly increased contrast. With a fiber of numerical aperture 0.22, we achieve a lateral resolution of 1.5μm, and an axial resolution of 12.7μm. The point-scanning rate is currently limited by our spatial light modulator (20Hz).

---

### References and links

1. C. J. R. Sheppard and A. Choudhury, "Image formation in the scanning microscope," Opt. Acta Int. J. Opt. **24**, 1051–1073 (1977).
2. T. Wilson, Ed., *Confocal Microscopy* (Academic, London, 1990).
3. P. Delaney and M. Harris, "Fiber-optics in scanning optical microscopy," in *Handbook of Biological Confocal Microscopy*, 3rd ed. (Springer US, 2006), pp. 508-515.
4. J. M. Jabbour, M. A. Saldua, J. N. Bixler, and K. C. Maitland, "Confocal endomicroscopy: instrumentation and medical applications," Ann. Biomed. Eng. **40**, 378–397 (2011).
5. A. F. Gmitro and D. Aziz, "Confocal microscopy through a fiber-optic imaging bundle," Opt. Lett. **18**, 565–567 (1993).
6. J. Knittel, L. Schnieder, G. Buess, B. Messerschmidt, and T. Possner, "Endoscope-compatible confocal microscope using a gradient index-lens system," Opt. Commun. **188**, 267–273 (2001).
7. Mauna Kea Technologies, "Cellvizio® Lab Technical Sheet," http://www.landing.maunakeatech.com/download-cellvizio-lab-technical-sheet.
8. J. T. Liu, M. J. Mandella, H. Ra, L. K. Wong, O. Solgaard, G. S. Kino, W. Piyawattanametha, C. H. Contag, and T. D. Wang, "Miniature near-infrared dual-axes confocal microscope utilizing a two-dimensional microelectromechanical systems scanner," Opt. Lett. **32**, 256–258 (2007).
9. H.-J. Shin, M. C. Pierce, D. Lee, H. Ra, O. Solgaard, and R. Richards-Kortum, "Fiber-optic confocal microscope using a MEMS scanner and miniature objective lens," Opt. Express **15**, 9113–9122 (2007).
10. C. J. Engelbrecht, R. S. Johnston, E. J. Seibel, and F. Helmchen, "Ultra-compact fiber-optic two-photon microscope for functional fluorescence imaging in vivo," Opt. Express **16**, 5556–5564 (2008).

11. B. H. W. Hendriks, W. C. J. Bierhoff, J. J. L. Horikx, A. E. Desjardins, C. A. Hezemans, G. W. 't Hooft, G. W. Lucassen, and N. Mihajlovic, "High-resolution resonant and nonresonant fiber-scanning confocal microscope," J. Biomed. Opt. **16**, 026007–026007–8 (2011).
12. Optiscan Pty. Ltd., "OptiScan FIVE 1 Product Brochure," http://www.optiscan.com/Products/FIVE1_Brochure.pdf.
13. R. Di Leonardo and S. Bianchi, "Hologram transmission through multi-mode optical fibers," Opt. Express **19**, 247–254 (2011).
14. A. M. Caravaca-Aguirre, E. Niv, D. B. Conkey, and R. Piestun, "Real-time resilient focusing through a bending multimode fiber," Opt. Express **21**, 12881–12887 (2013).
15. S. Rosen, D. Gilboa, O. Katz, and Y. Silberberg, "Focusing and scanning through flexible multimode fibers without access to the distal end," http://arxiv.org/abs/1506.08586 (2015).
16. I. N. Papadopoulos, S. Farahi, C. Moser, and D. Psaltis, "Focusing and scanning light through a multimode optical fiber using digital phase conjugation," Opt. Express **20**, 10583–10590 (2012).
17. I. N. Papadopoulos, S. Farahi, C. Moser, and D. Psaltis, "High-resolution, lensless endoscope based on digital scanning through a multimode optical fiber," Biomed. Opt. Express **4**, 260–270 (2013).
18. E. E. Morales-Delgado, S. Farahi, I. N. Papadopoulos, D. Psaltis, and C. Moser, "Delivery of focused short pulses through a multimode fiber," Opt. Express **23**, 9109–9120 (2015).
19. T. Čižmár and K. Dholakia, "Exploiting multimode waveguides for pure fibre-based imaging," Nat. Commun. **3**, 1027 (2012).
20. S. Bianchi and R. Di Leonardo, "A multi-mode fiber probe for holographic micromanipulation and microscopy," Lab. Chip **12**, 635–639 (2012).
21. Y. Choi, C. Yoon, M. Kim, T. D. Yang, C. Fang-Yen, R. R. Dasari, K. J. Lee, and W. Choi, "Scanner-free and wide-field endoscopic imaging by using a single multimode optical fiber," Phys. Rev. Lett. **109**, 203901 (2012).
22. M. Plöschner, T. Tyc, and T. Čižmár, "Seeing through chaos in multimode fibres," Nat. Photonics **9**, 529–535 (2015).
23. J. Carpenter, B. J. Eggleton, and J. Schröder, "110x110 optical mode transfer matrix inversion," Opt. Express **22**, 96–101 (2014).
24. D. Kim, J. Moon, M. Kim, T. D. Yang, J. Kim, E. Chung, and W. Choi, "Toward a miniature endomicroscope: pixelation-free and diffraction-limited imaging through a fiber bundle," Opt. Lett. **39**, 1921–1924 (2014).
25. A. S. Goy and D. Psaltis, "Digital confocal microscope," Opt. Express **20**, 22720–22727 (2012).
26. A. S. Goy, M. Unser, and D. Psaltis, "Multiple contrast metrics from the measurements of a digital confocal microscope," Biomed. Opt. Express **4**, 1091–1103 (2013).
27. M. Jang, H. Ruan, H. Zhou, B. Judkewitz, and C. Yang, "Method for auto-alignment of digital optical phase conjugation systems based on digital propagation," Opt. Express **22**, 14054–14071 (2014).
28. S. M. Popoff, G. Lerosey, R. Carminati, M. Fink, A. C. Boccara, and S. Gigan, "Measuring the transmission matrix in optics: an approach to the study and control of light propagation in disordered media," Phys. Rev. Lett. **104**, 100601 (2010).
29. S. M. Popoff, G. Lerosey, M. Fink, A. C. Boccara, and S. Gigan, "Controlling light through optical disordered media: transmission matrix approach," New J. Phys. **13**, 123021 (2011).
30. M. Plöschner, B. Straka, K. Dholakia, and T. Čižmár, "GPU accelerated toolbox for real-time beam-shaping in multimode fibres," Opt. Express **22**, 2933–2947 (2014).
31. M. Plöschner and T. Cizmár, "Compact multimode fiber beam-shaping system based on GPU accelerated digital holography," Opt. Lett. **40**, 197–200 (2015).
32. A. Yariv, "Fundamental media considerations for the propagation of phase-conjugate waves," Opt. Lett. **16**, 1376–1378 (1991).
33. T. Čižmár, M. Mazilu, and K. Dholakia, "In situ wavefront correction and its application to micromanipulation," Nat. Photonics **4**, 388–394 (2010).
34. S. A. Goorden, J. Bertolotti, and A. P. Mosk, "Superpixel-based spatial amplitude and phase modulation using a digital micromirror device," Opt. Express **22**, 17999–18009 (2014).
35. E. Sánchez-Ortiga, A. Doblas, G. Saavedra, M. Martínez-Corral, and J. Garcia-Sucerquia, "Off-axis digital holographic microscopy: practical design parameters for operating at diffraction limit," Appl. Opt. **53**, 2058–2066 (2014).
36. G. Barbastathis, M. Balberg, and D. J. Brady, "Confocal microscopy with a volume holographic filter," Opt. Lett. **24**, 811–813 (1999).
37. S. Farahi, D. Ziegler, I. N. Papadopoulos, D. Psaltis, and C. Moser, "Dynamic bending compensation while focusing through a multimode fiber," Opt. Express **21**, 22504–22514 (2013).
38. A. M. Caravaca Aguirre and R. Piestun, "Robustness of multimode fiber focusing through wavefront shaping," in Latin America Optics and Photonics Conference, OSA Technical Digest (online) (Optical Society of America, 2014), p. LTh4A.23.
39. B. Redding, S. M. Popoff, and H. Cao, "All-fiber spectrometer based on speckle pattern reconstruction," Opt. Express **21**, 6584–6600 (2013).

40. S. González and Z. Tannous, "Real-time, in vivo confocal reflectance microscopy of basal cell carcinoma," J. Am. Acad. Dermatol. **47**, 869–874 (2002).
41. C. Liang, M. Descour, K.-B. Sung, and R. Richards-Kortum, "Fiber confocal reflectance microscope (FCRM) for in-vivo imaging," Opt. Express **9**, 821–830 (2001).
42. D. Andreoli, G. Volpe, S. Popoff, O. Katz, S. Grésillon, and S. Gigan, "Deterministic control of broadband light through a multiply scattering medium via the multispectral transmission matrix," Sci. Rep. **5**, (2015).
43. I. M. Vellekoop, M. Cui, and C. Yang, "Digital optical phase conjugation of fluorescence in turbid tissue," Appl. Phys. Lett. **101**, 081108 (2012).
44. P. J. Schreier and L. L. Scharf, "Correlation analysis," in *Statistical Signal Processing of Complex-Valued Data: The Theory of Improper and Noncircular Signals* (Cambridge University, 2010), pp. 85-88.

## 1. Introduction

*1.1 Fiber-based confocal endoscopes*

Confocal microscopy is an important tool in biological imaging, because it substantially improves the contrast of images compared to wide field microscopy, and it allows depth-sectioning [1, 2]. In essence, the confocal microscope is based on a double filtering operation: a certain volume inside the sample is selectively illuminated by a focused beam, and light originating from this focal volume is selectively observed using a pinhole in the detection pathway. The pinhole is located in a plane conjugated with the focal plane, and suppresses light originating from any location other than the focal volume. With this method, a point of a sample can be probed with higher contrast with respect to its surroundings. Images are built by scanning the probed focal volume inside the sample.

In typical biological media, confocal microscopy allows us to obtain clear, background-free images only up to a certain point. Indeed, when focusing at a depth larger than the scattering mean free path, photons on the illumination path are scattered away before they can reach the focal volume. On the detection side, they are diverted from the detection path and blocked by the pinhole. The resulting loss in sensitivity ultimately limits the confocal imaging depth to the superficial layers of the tissue.

To image biological structures that are located deep in tissue, fiber-based endoscopes can provide a minimally invasive solution. The existing confocal fiber endoscopes can be divided into two categories: fiber bundle systems and distal scanning systems [3, 4].

In fiber bundle systems, a coherent fiber bundle relays the spots created by a conventional confocal microscope to the distal facet of the bundle. The plane of imaging is either the distal facet of the bundle itself (the sample must then be placed in contact with this surface), or an extra lens (e.g. a GRIN rod lens) can be attached to the distal tip of the bundle in order to move the focal plane some distance away from the tip [5–7]. This arrangement allows for thin endoscopes (300μm – 1mm), but the resolution is limited because of the required inter-core spacing of the bundle, which is in general 3μm or more. A magnifying element can be used at the tip to improve the effective resolution, but in that case diffraction-limited spots may overfill the individual cores of the bundle, decreasing the system's collection efficiency [3]. In addition, magnification reduces the field of view below the probe's size.

Another approach is to add a miniature scanning mechanism at the tip of a single-mode fiber. For example, a MEMS scanner can be used to scan the light beam [8, 9] or the fiber tip itself can be scanned [10–12]. Such devices can reach diffraction-limited resolution, but have large probes of several millimeters.

*1.2 Multimode fiber imaging*

Recently, multimode fibers have been shown to be an interesting alternative for endoscopic applications thanks to their ability to guide many independent spatial modes of light within a very small cross-sectional diameter, down to 100μm. The multiple modes allow these fibers to transmit images composed of multiple pixels with diffraction-limited resolution, whereas single-mode fibers can only transmit light with a Bessel intensity profile and bundles of single-mode fibers are limited in resolution by the required inter-core spacing between the fibers.

The main difficulty in exploiting fiber modes for imaging is that different modes travel with different propagation constants and modes can also couple to each other. Concretely, this means that an image fed into one side of a multimode optical fiber will not retain its shape as it propagates to the other side of the fiber. While the information about the image becomes scrambled in this process, it is however not destroyed. The image can be reconstructed given the knowledge of the propagation characteristics of light inside the fiber.

Several techniques have been developed to undo the effects of modal scrambling in multimode fibers. These techniques record the association between the images at the input of the fiber and the scrambled patterns at the output during a calibration phase. Optimization techniques [13–15] iteratively find the output pattern associated with each image from a predetermined set of inputs. In digital optical phase conjugation [16–18] such output patterns are recorded with a holographic acquisition. The transmission matrix method [19–23] captures the propagation characteristics of the fiber in a matrix linking the input field with the output field. Once the fiber is calibrated, a light modulator is used to shape the input wavefront sent to the fiber, so that it creates spots or other known patterns at the opposite end; these spots or patterns are used to probe the sample. The scanning rate is currently in the kilohertz range in fastest implementations [19, 21, 24], yielding an effective frame rate of about 1Hz for high-resolution scans. Most implementations to date were based on narrowband lasers, but extension to broadband pulsed lasers is underway [15, 18].

Ideally, the same multimode fiber should provide illumination to the location of interest, as well as guide the resulting signal back to a detector. Currently, two such imaging mechanisms have been successfully implemented: non-confocal fluorescence imaging [17, 19, 20, 24] and wide field reflection imaging [21, 22]. In these demonstrations, modal scrambling was compensated either on the way in or on the way out of the fiber, but not both at the same time. For the fluorescence imaging experiments, the input light was modulated to focus spots on a sample, and the returning fluorescence signal was isolated by means of a dichroic mirror. In the wide field reflection experiments, the sample was illuminated by a random set of speckle fields, and only the returning light was decoded to retrieve spatial information.

*1.3 Confocal microscopy through multimode fibers*

Here we propose a digital implementation of confocal microscopy combined with multimode fiber imaging. For this, the modal distortions need to be compensated both ways in order to select a particular focal volume during both excitation and detection. In the digital variant of confocal microscopy [25], the light returning from the sample is recorded with digital holography in an intermediate plane, instead of being filtered by a physical pinhole in a conjugate plane. The field is then digitally propagated up to a virtual conjugate plane, where it forms a focus. The digitally focused field can finally be filtered with a virtual pinhole mask, making the detection spatially selective as in classical confocal microscopy. The digital detection of the optical fields provides a large flexibility in the signal processing, allowing for example the dynamic adjustment of the pinhole size as well as the measurement of new contrast metrics such as the focal phase or the focal width [26]. In our case, it also allows to correct for the distortions due to the fiber before filtering with a pinhole.

Practically, we use a multimode fiber to guide light to and from the location of interest of a sample, and we implement reflection-mode (non-fluorescent) digital confocal detection at the multimode fiber's tip (see Fig. 1). Prior to the experiment, we measure the transmission matrix (TM) of a multimode fiber and use it to project arbitrary illumination patterns, as well as decode the fields propagating in the reverse direction through the same multimode fiber. Then, we implement the digital filtering required for confocal microscopy. The purpose is to increase the imaging contrast in spot-scanned images, which is important for applications such as imaging inside scattering tissues. A correlation-based filtering technique is also introduced, which offers similar performance for a significantly reduced computational cost.

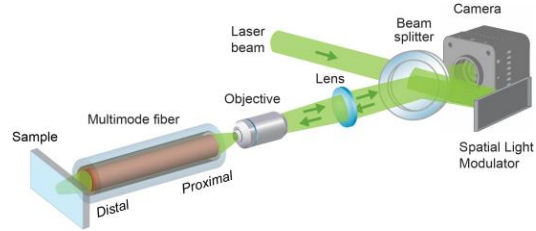

Fig. 1. Overview of the multimode fiber confocal system. A reference laser beam is reflected off a spatial light modulator (SLM). The phase modulated light beam is injected into the fiber in order to produce the desired excitation beam at the sample plane. The field collected back through the fiber is digitally processed in order to render a confocal image of the sample.

## 2. Methods

### 2.1. Imaging setup

The output of a diode-pumped solid-state laser at 532nm (CNI MSL-FN-532-100mW) is spatially filtered and collimated to form a plane wave reference beam. After being split by a beamsplitter, the plane waves travel to each side of the multimode fiber (Thorlabs M43L01, Ø105µm core, 0.22 NA, FC-APC). Off-axis holography is used to measure the fields coming out of the fiber. On each side, the fiber facet is first magnified with a microscope objective (Newport MV-40x) and imaged via a lens (Thorlabs AC254-250-A-ML, f = 250mm) onto a camera sensor (PhotonFocus MV1-D1312(IE)-G2-100), where the light field is interfered with the reference plane wave. This is detailed in Fig. 2. The angle between the reference beam and the object beam for off-axis holography is approximately 1.5°.

We wish to transmit images in both directions through the multimode fiber, and to avoid confusion we will now designate the side of the fiber with the spatial light modulator as the "proximal side". This is where we control the illumination and perform the confocal detection. The other side is called the "distal side". In the distal side, the holographic acquisition system is used only for calibration. It is the side where the sample is located, and during imaging it is devoid of any hardware besides the fiber itself. In the proximal side, a spatial light modulator (HoloEye Pluto) is used to illuminate the fiber with controlled patterns at a maximal rate of 20Hz.

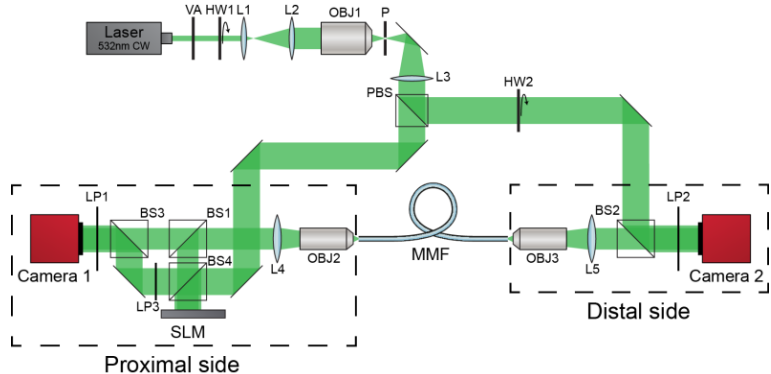

Fig. 2. Detailed diagram of the experimental setup. VA: variable attenuator; HW1, HW2: half-wave plate; L1: f=30mm lens; L2: f=75mm lens; L3: f=200mm lens; OBJ1: infinity corrected 10x microscope objective; P: 10µm pinhole; L3: f=200mm lens; PBS: polarizing beamsplitter; BS: beamsplitter; LP: linear polarizer; SLM: spatial light modulator; L4, L5: f=250mm lenses; OBJ2, OBJ3: 40x microscope objective; MMF: multimode fiber.

## 2.2. Transmission matrix

The first step in controlling the modes of a multimode fiber is to determine how they are transformed between the input and the output of the fiber i.e. measuring its transmission matrix (TM). This can be done experimentally by applying the modes one by one to the input of the fiber with a spatial light modulator (SLM), and recording the corresponding output fields holographically. Each of these measurements yields one column of the transmission matrix. Assuming a complete set of modes is sampled, the transmission matrix can be used to predict any future output by linear combination of the known input-output measurements.

It is possible to use the theoretical modes of the fiber as input basis for this procedure [22, 23], but any other set of linearly independent input patterns is equally suitable as long as it can properly describe fields entering and exiting the fiber. We chose a basis of plane waves with varying spatial frequencies (i.e. varying angles with respect to the optical axis), because plane waves can accurately be displayed on a phase-only SLM and no light is lost at angles outside the numerical aperture of the fiber (unless those angles are explicitly probed). The relationship between the plane wave basis and the physical pixel basis of the SLM is simply a Fourier transform. This is shown in Fig. 3(a).

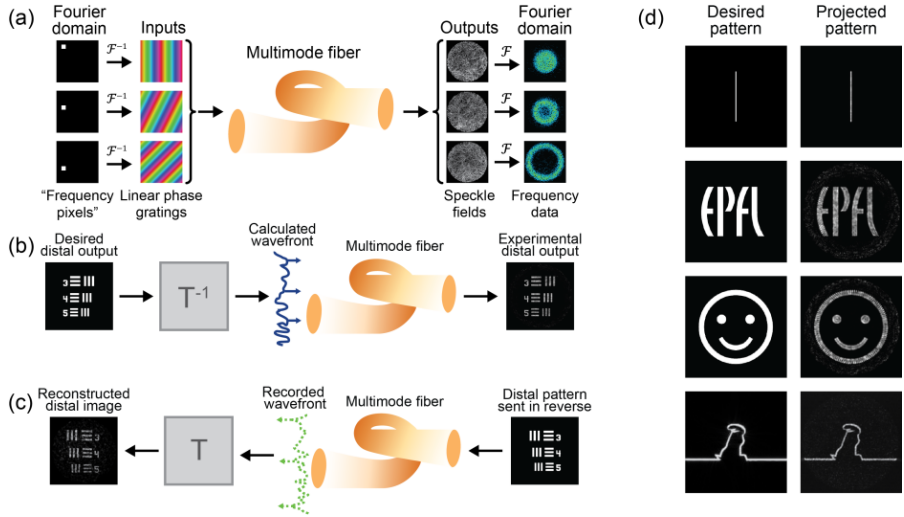

Fig. 3. (a) The transmission matrix is measured using a basis of plane waves, which are equivalent to pixels in the Fourier domain. Each 'Fourier pixel' is successively turned on and off, and the corresponding output pattern is recorded. Each output pattern forms one column of the transmission matrix. (b) Pattern projection: based on a digital image of part of a USAF1951 pattern, a wavefront is calculated using the inverse transmission matrix. This wavefront is then generated by the SLM and sent through the fiber from the proximal end. On the distal of the fiber, the pattern emerges. (c) Reverse image transmission: a part of a USAF1951 target is illuminated from behind by a collimated beam and imaged onto the distal facet of the fiber. The wavefront generated in the proximal end is recorded, and decoded using the transmission matrix. In this way, the pattern can be reconstructed digitally. (d) Experimental results for image transmission. The last row is a snapshot from an animated cartoon ('La Linea', 1971, Osvaldo Cavandoli); see also the associated Visualization 1.

## 2.3. Camera-SLM alignment

To transmit images in the reverse direction (i.e. from distal to proximal), as needed for confocal filtering, the field should to be recorded holographically at the proximal end and then reconstructed using the transmission matrix. Here lies a significant practical challenge: the transmission matrix is measured using the SLM, but the field must be recorded separately with a camera. For an accurate reconstruction, the camera must record the field exactly as it exists at the position of the SLM. This is possible by placing the SLM and the camera in equivalent planes behind a beamsplitter, as shown in Fig. 1. The two devices must be aligned precisely in position and in angle, and should ideally have the same pixel pitch. In our experiments, the tolerances for the SLM position leading to a 5% change in reconstructed spot intensity were 15µm of lateral translation and 4 arcsec of rotation perpendicularly to the optical axis, and 1.2cm for a translation and 0.9° for a rotation along the optical axis. To reach the required precision quickly and easily, we used a digital registration approach. The fields captured with the camera were interpolated, displaced, and tilted so as to match the coordinate system of the SLM. The alignment parameters can be tuned only once and stay stable until either component is moved. Similar strategies can be found in literature for the alignment of digital phase conjugation mirrors [27].

*2.4 Bidirectional image transmission*

In order to calculate which proximal field will create a given pattern at the distal side of the fiber, the transmission matrix needs to be inverted. Because of measurement noise, a regularized inversion scheme is necessary. We used Tikhonov inversion, which has been successfully applied in the context of scattering media before [28, 29]:

$$T^{-1} = V S^+ U^H \qquad (1)$$

where $USV^H$ is the singular value decomposition (SVD) of the transmission matrix $T$. $U^H$ and $V^H$ denote the Hermitian transposes of matrices $U$ and $V$. The singular values $\sigma_i$ of the matrix $T$ are located on the diagonal of $S$. In $S^+$ these values are spectrally filtered by $\sigma_i / (\sigma_i^2 + \lambda^2)$, as required for Tikhonov inversion. The regularization parameter $\lambda$ was chosen as 10% of the greatest singular value $\sigma_1$.

Once the inverse is calculated, any illumination pattern can be displayed dynamically at the distal end of the multimode fiber, as shown in Figs. 3(b) and 3(d), and Visualization 1. Thanks to the use of phase tracking during the measurement of the transmission matrix and Gerchberg-Saxton encoding of the modulated wavefronts (further explained in the Appendix), the patterns do not suffer from interference artefacts [30, 31]. We obtain a linear correlation of over 95% between the experimental intensity patterns and the desired intensity patterns.

To transmit images in the reverse direction through the system (i.e. from the distal to the proximal side), we record the field in the proximal end with a single holographic acquisition and decode it using the transmission matrix. This yields the field at the distal end of the fiber (Fig. 3(c)). The alignment of the SLM and the holographic acquisition in the proximal end is critical for the successful reconstruction of the distal field, as explained before.

*2.5 Digital confocal microscopy*

For the confocal scanning, we first put the appropriate pattern on the proximal SLM in order to generate an excitation spot at a distance of approximately 100µm in front of the distal fiber facet. This spot interacts with the sample at that location, and the reflected and backscattered light is collected back through the fiber. The field is then recorded holographically at the proximal side. One such measurement is performed for each position of the sample.

Three ways of processing the acquired data were tested. In the first method we simply integrate the total intensity of the proximally recorded field. This serves as a reference image, showing the contrast that would be obtained if the returning light were measured with a bucket photodetector without any further processing.

The second method is the digital confocal method. Here, we use the transmission matrix to virtually propagate the backscattered field back through the fiber, and reconstruct it as it existed at the position of the sample. There, we apply a digital pinhole mask that suppresses all light contributions except those found within a radius of 1µm of the position of the excitation spot. Note that the Rayleigh radius for this wavelength (532nm) and fiber NA (0.22) is 1.5µm. The light energy that remains after filtering with the digital pinhole is integrated, and this value forms one pixel of the final image. This filtering scheme is illustrated in Fig. 4(a).

We refer to the last method as the correlation method, and it is based on a different filtering principle. Consider the field that is sent from the proximal end in order to create a focus spot at the distal end of the fiber. The light that originates from that same spot at the distal end and carrying the sample information propagates back through the fiber towards the proximal side, where it should lead to a similar field as we used for excitation (neglecting losses), simply because of the reversibility of light propagation. The phase conjugation literature [16, 32] provides formal and experimental proof of this principle. Any contribution of light not originating from the focal point should, on the contrary, lead to a proximal field that is uncorrelated with the excitation field due to the randomizing nature of modal scrambling. Therefore, the distal spot intensity can be estimated simply by calculating the linear projection (or correlation) of the returning field with respect to the excitation field, as shown in Fig. 4(b). This operation is done for each scanning spot and the image is constructed pixel by pixel.

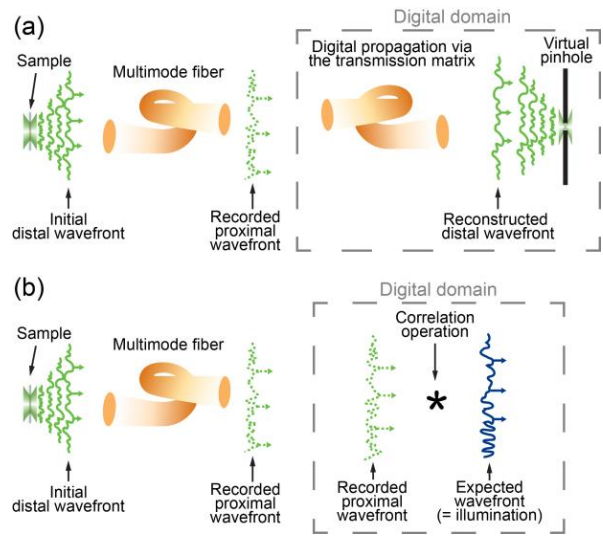

Fig. 4. (a) In the digital confocal method, the recorded field is virtually propagated back to the position of the sample via the transmission matrix. There, it is filtered with a pinhole mask. (b) In the correlation method, the returning field is correlated with the illumination field.

## 3. Results

Our first set of experiments consisted of imaging of a human epithelial cell dried on a microscope cover glass. The results are shown in Figs. 5(a)-5(c). The image area is 81µm by 76µm, and the step size is 1.1µm. A control image made in white light transmission is shown in Fig. 5(d).

A similar experiment was made for polystyrene beads spread on the surface of a cover glass. These results are shown in Figs. 5(e)-5(g), with a control image in Fig. 5(h). Here the area is 22.5µm by 22.5µm, and the step size is 0.55µm. To have an estimate for the resolution of our system, we calculated the full width at half maximum of one of the reconstructed spots in the digital confocal image (Fig. 5(f)), which is 1.5µm.

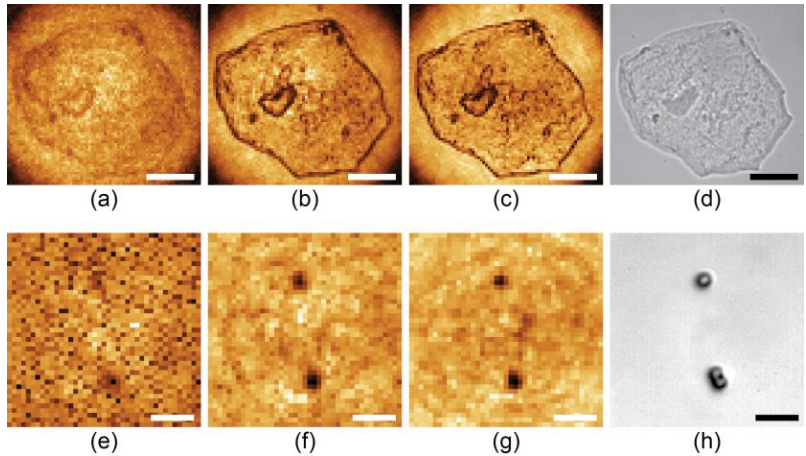

Fig. 5. (a-d) Microscopic image of a human epithelial cell reconstructed using (a) the total intensity method, (b) the digital confocal method, (c) the correlation method. (d) Control image taken in transmission, i.e. observed from the distal end with a camera using white-light illumination. (e-h) 1µm polystyrene beads imaged with (e) the total intensity method, (f) the confocal method, (g) the correlation method, (h) control. (a-d) Scale bar is 20µm. (e-h) Scale bar is 5µm.

In a second experiment, we made a transversal scan (z-scan) of a cover glass, as sketched in Fig. 6(e). This is to illustrate the sectioning capability that we can obtain using the proposed filtering techniques. The results are shown in Figs. 6(a)-(c). A control image is shown in Fig. 6(d); it was taken on a commercial laser-scanning confocal microscope (Zeiss LSM 710) with an NA 0.3 objective. The average full width at half maximum of the interface is 12.7µm in the digital confocal image, 15.8µm in the correlation image, and 10µm in the control image. The ratio of the coverslip signal to the average background intensity between the interfaces is 22.5:1 in the digital confocal image (Fig. 6(b)), 8.4:1 in the correlation image (Fig. 6(c)) and 270:1 in the control image (Fig. 6(d)).

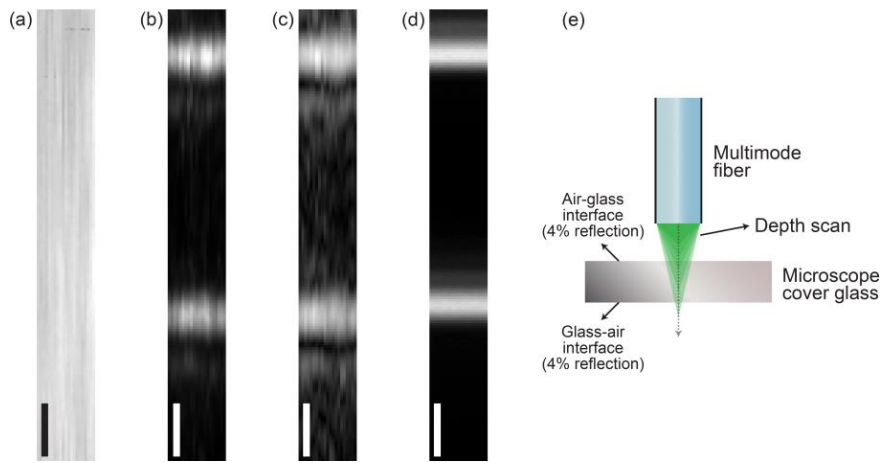

Fig. 6. Transversal scans of a coverslip with the (a) total intensity method, (b) digital confocal method, (c) correlation method and (d) control image taken with a commercial confocal microscope. The scale bars represents 20µm of distance in air. Note that the thickness of the coverslip is approximately 150µm, but due to refraction it appears thinner in these images. The vertical axis is perpendicular to the coverslip, and the horizontal axis represents a lateral scan. (e) Schematic description of the experiment.

## 4. Discussion

*4.1 Contrast enhancement, sectioning and image quality*

The comparison of the various methods in Fig. 5 reveals that a significant increase in image contrast is achieved when filtering the backscattered light, versus the case where the whole field is integrated. By digitally implementing spatial selectivity in the detection, we were able to clearly distinguish the walls and the nucleus of an epithelial cell in Figs. 5(b) and 5(c). Also in the case of polystyrene beads, the filtering scheme was useful. With this sample, we recorded an intensity image with very little contrast in Fig. 5(e), but the beads appear clearly on the confocal and correlation images Figs. 5(f) and 5(g).

Similarly, the depth scans of Fig. 6 show reflective interfaces could not be resolved by simply recording the total backscattered intensity (Fig. 6(a)), but they were made visible by the proposed filtering schemes (Figs. 6(b) and 6(c)). Due to the limited numerical aperture of the fiber (NA 0.22), the axial resolution is relatively low in Figs. 6(b) and 6(c). The numerical aperture explains part of the difference between these images and the control image from a traditional microscope (NA 0.3). Note that the transmission matrix method is general and can be used with any type of fiber. Therefore, the steps outlined in this manuscript can be extended to fibers with a higher numerical aperture or a larger core; however, this implies that a greater number of modes need to be sampled during calibration, and with a slow modulator it is preferable to keep this number low (the calibration currently takes 10min in our implementation).

Other factors play a role as well in determining the image quality obtained with our approach. In the experiments presented here, we illuminated and recorded only one polarization of the light going through the fiber, for experimental simplicity. Since the fiber acts as a depolarizing medium for linear polarization, half of the light is lost each way. Polarization multiplexing techniques [22, 23] may improve the sensitivity by allowing to process all of the light travelling through the fiber. An added benefit of polarization multiplexing would be the capability to do confocal polarization microscopy.

Finally, most phase-only spatial light modulators are known to cause aberrations due to the fact that their surfaces are not perfectly flat. This induces a systematic error in the measurement of the transmission matrix. Because the same aberration is not present on the camera used for recording backscattered field, it is not possible to perfectly reconstruct the distal field from the proximally measured data. One possible solution is to use a modulator that is flat or corrected for such errors, or measure the deformation experimentally and correct for it [33].

*4.2 Speed*

The experiments presented here are currently limited in speed by our modulator. With a point-scanning rate of 20Hz, the measurement shown in Figs. 5(a)-5(c) took 4min 15s to acquire, Figs. 5(e)-5(f) took 1min 24s, and Figs. 6(a)-6(c) took 3min. Faster modulators can be used, such as digital micromirror devices or a combination of an acousto-optic deflector with a spatial light modulators. These have been shown to work for similar applications [14, 19, 24, 34], and reach speeds over 20kHz.

The next limiting factors would be the speed of the acquisition (i.e. the frame rate of the camera), and ultimately the computational load of reconstructing holograms. We use digital off-axis holography here, and with this method the speed of reconstruction is mainly determined by speed of the necessary Fourier transform. On a computer with an Intel Xeon E5-2620, using the FFTW library, we were able to process holograms of 800 by 800 pixels at a speed of 400 frames per second. Note that in the digital confocal method, two Fourier transforms are required: one to reconstruct the off-axis hologram captured in the proximal side, and one to reconstruct the distal field from the unscrambled Fourier coefficients calculated with the transmission matrix. With the correlation method, only the first transform is needed (for the holographic reconstruction).

The processing speed can be increased by making lower-resolution holograms. A lower-resolution means that the field of view has to be reduced, and/or the magnification of the optical detection system (OBJ2, L4, OBJ3 and L5 in Fig. 2) should be reduced, leading to a smaller spatial frequency bandwidth [35]. The resolution of 800 by 800 pixels that we used is enough for fibers with a V number up to 350, e.g. a fiber with a core of 105μm and NA 0.56 or a fiber with NA 0.22 and a core of 270μm at 532nm.

*4.3 Comparison of the digital confocal and the correlation method*

In effect, the pinhole method performs the same operation as a classical confocal microscope, while the correlation method acts more like a matched filter [36] measuring the amount of backscattered light bearing the same signature as the excitation light. The correlation method has a lower computational cost, because we do not need to transform the proximal field and reconstruct the distal field. However, there is also less flexibility in the signal processing, since the pinhole size cannot be adjusted and the reconstructed spots are not available for further analysis.

As opposed to the digital confocal method, the correlation method can be completely hardware-implemented by letting the backscattered field reflect on the SLM. This field will then be demodulated by the phase pattern currently being displayed. In other words, the backscattered field (the field to be filtered) will be multiplied by the illumination pattern (the field we wish to correlate with). After this operation, the light can simply be focused through a lens to obtain the Fourier transform, and a pinhole can be used to extract the DC-term of the resulting field. In this case, the acquisition speed would only be limited by the modulator.

*4.4 Bending and stability*

The proposed methods depend on the characterization of the fiber by the transmission matrix, and this transmission matrix changes depending on the bending state of the fiber. While there is a certain limited tolerance to bending [21, 37, 38], for practical applications it may be preferable to use a fiber immobilized inside a needle [17], as a rigid endoscope. The small outer diameter (125 - 300μm) of multimode fibers is compatible with some of the thinnest needle gauges, so this constitutes a minimally invasive method for deep-tissue microscopy.

Other proposals in literature that address the problem of bending include using a semi-rigid probe, with a calibration stored for a discrete set of bending states [37], or compensating bending in real-time with a fast feedback system [14]. By using two-photon fluorescence as a feedback signal and exploiting the structure of light patterns in graded-index fibers, it is possible to obtain the calibration of the fiber without access to the distal end [15].

Recently, it was demonstrated that the transmission matrix of a fiber can be calculated instead of being measured [22]. It is also possible to calculate the matrix for different bending states of the fiber. This study suggests that it may be possible to compensate for the bending of the fiber by recalculating the matrix in real-time. The images acquired through the fiber endoscope could be used as feedback signal in order to estimate the bending state.

Another important point with regard to the proposed applications is the temperature stability of the transmission matrix. According to previous results in literature [39], the temperature variation that is necessary to decorrelate a speckle pattern through a 1m long fiber is 8°C, and this scales inversely with fiber length. Therefore, it may be necessary to calibrate the fiber at the temperature of the body, but there is otherwise enough temperature margin for most endoscopic applications.

*4.5 Fluorescence*

Here, we showed results for reflection-mode confocal operation. This has the advantage for in-vivo operation that no fluorescent probes need to be injected to the area of interest before it can be imaged, i.e. the technique works label-free [40, 41]. If a label is desired, for example to target specific parts of a tissue, one should use scattering probes such as nanoparticles.

Since confocal microscopy is often used in biology to image fluorescent specimens, we briefly discuss whether the proposed methods can be extended to this case. For imaging fluorescence, the transmission matrix must in principle be known for both the excitation and the emission wavelength. With this information, the correlation method could be implemented as follows: fluorescence emission could be spectrally filtered to yield a speckle pattern, and this speckle pattern could be correlated with the pattern expected for fluorescence emission from the excited spot. Multispectral transmission matrices have been studied before in the context of scattering media [42]. The fluorescence bandwidth that could be obtained with such a technique depends on the spectral decorrelation width, which is inversely proportional to the length of the fiber [39]. The digital confocal method relies on holographic detection. Since there is no coherent reference available in the case of fluorescence, a reference-free method of recovering the phase information would be needed to apply this method [43].

## 5. Conclusion

Our experiments show that the principle of confocal filtering is broadly applicable, even in cases where the light paths towards the focal volume are severely distorted. The schemes presented here can be generalized to any system where the distortion is described by a transmission matrix, e.g. also in scattering media [28].

In the context of biomedical imaging, the multimode fiber can be calibrated outside the tissue of interest, and then inserted at another location (i.e. inside the tissue) for imaging. The proposed system does not have any distal scanning optics, and the probe diameter can therefore be as thin as the fiber itself. The focal plane can be chosen dynamically by appropriate modulation from the proximal side.

We proposed two conceptually different ways of obtaining a confocal filtering effect via multimode fibers. This has potential applications in the endoscopic high-contrast imaging of cells, either label-free or with scattering probes such as nanoparticles.

## Appendix

### A.1 Phase tracking

Due to the limited rate at which input patterns can be applied with the spatial light modulator (20Hz), it is important to monitor the stability of the system over time while measuring the transmission matrix. The phase between the reference beam for holography and the object beam coming from the fiber is particularly important: if the columns of the transmission matrix are measured with a varying phase reference, these columns cannot be used together in a coherent fashion. Thus, we implemented a simple phase tracking scheme where, during the measurement of the transmission matrix, a reference input is sent to the fiber at constant time intervals. The phases of the output fields corresponding to these reference inputs are compared with each other over time. The comparison is done using the complex correlation coefficient, $\rho = \langle f, g \rangle / \|f\| \|g\|$, where $f$ and $g$ are two complex vectors to compare to each other. The magnitude of this number gives the degree of linear similarity between the two vectors. The phase encodes for the average phase rotation between them [44]. Any detected phase drift in the columns of the transmission matrix is corrected by interpolation. A typical trace of the phase drift over time can be found in Fig. 7. The magnitude of the correlation coefficient reveals that the field distribution does not change significantly over time. However, over a measurement period of 10 minutes, we observe a phase change of approximately 360° between the start and the end of the experiment.

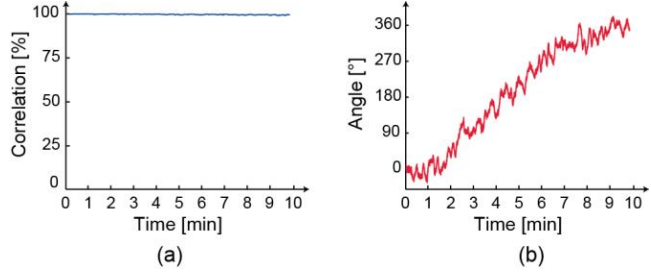

Fig. 7. Complex correlation coefficient calculated between the initial output field and the output field at all later times. (a) The magnitude of the correlation coefficient reveals that the field distribution does not change significantly over time. (b) The phase drift can be measured over time, and corrected by interpolation.

*A.2 Gerchberg-Saxton encoding*

A phase-only spatial light modulator is ideal to generate patterns such as plane waves of varying spatial frequencies with high efficiency. However, to create arbitrary patterns at the output of the multimode fiber, the required input patterns generally have both amplitude and phase variations. Modulating only the phase degrades the achievable signal-to-background ratio [29]. This can be acceptable when generating spots, because focusing all of the available light to only one point usually results in signal-to-background ratios in excess of 1000:1. To generate more general patterns such as shown in Fig. 3(b), a more optimal strategy is needed. For these patterns, we used the Gerchberg-Saxton algorithm [31], as explained in Fig. 8. Briefly, this algorithm finds a field satisfying constraints both in the spatial domain as well as in the Fourier domain. In the spatial domain, the modulator constrained the field to have constant amplitude. In the Fourier domain, we held constant the Fourier coefficients corresponding to spatial frequencies within the numerical aperture of the fiber. These coefficients represent the field that will actually be coupled to the fiber, and we set it equal to the field we calculate from the transmission matrix. The remaining coefficients (not coupled to the fiber) are chosen freely by the algorithm to respect the constraints. With this algorithm, it is possible to calculate phase patterns that closely approximate full phase and amplitude modulation in practice.

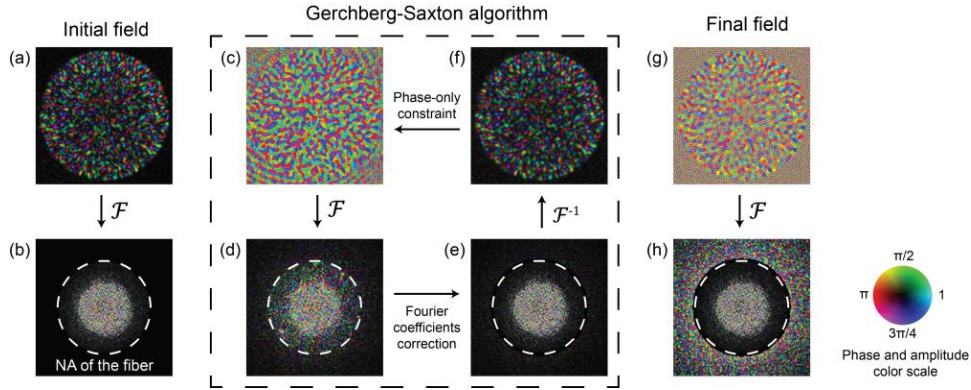

Fig. 8: Gerchberg-Saxton algorithm for phase-only encoding. (a) The initial field. (b) The Fourier transform of the initial field. The Fourier transform is a decomposition into plane waves with varying angles with the optical axis. In the Fourier domain, we can therefore distinguish two zones with respect to the fiber: there are coefficients corresponding to plane waves within the numerical aperture (NA) of the fiber, and plane waves outside the NA of the fiber. The boundary between both zones is shown with a dashed line. (c) In the first step of the algorithm, we make the field phase-only in the spatial domain by setting the amplitude of each pixel to 1. (d) In the Fourier transform of (c), we observe that the phase-only operation has created additional components outside the NA of the fiber, and has also distorted the components within the NA of the fiber. (e) The second step of the algorithm is to correct Fourier coefficients that were distorted by the phase-only operation. Here, we simply replace these coefficients with the undistorted coefficients of the initial field within the NA of the fiber. The coefficients outside the NA of the fiber are left as calculated by the algorithm; their value can be freely modified since this corresponds to light that will be attenuated inside the fiber. (f) In the spatial domain, the correction of the Fourier coefficients has recreated a non-constant distribution of amplitude. The process (c) to (f) is therefore repeated for several iterations. (g) After 50 iterations, a phase-only field is obtained. (h) The final field has the desired Fourier components within the numerical aperture of the fiber. Note that in our experiments, the fields (a), (c), (f), (g) and their Fourier transforms (b), (d), (e), (f) have a resolution of 800 by 800 pixels, but for clarity only the central 125 by 125 pixels of the Fourier transforms are shown in (b), (d), (e), (h).

## Acknowledgment

This study was funded in part by the Swiss National Science Foundation (grant no. 200021_160113/1, project MuxWave).